\documentclass[prl,twocolumn,showpacs]{revtex4}

\usepackage{epsfig}
\usepackage[centertags]{amsmath}
\usepackage{graphicx}

\begin{document}

\def\Ep#1{Eq.\ (\ref{#1})}
\def\Eqs#1{Eqs.\ (\ref{#1})}
\def\EQN#1{\label{#1}}
\def\ket{\rangle}
\def\bra{\langle}
\newcommand{\beqa}{\begin{eqnarray}}
\newcommand{\eeqa}{\end{eqnarray}}
\def\eps{\epsilon}
\def\xit{{\tilde \xi}}

\title{Bound states in continuum in an electron waveguide  }

\author{Kyungsun Na}
\affiliation{Center for Studies in Statistical Mechanics and Complex Systems,
\\ The University of Texas at Austin, Austin, TX 78712 USA }

\author{Gonzalo Ordonez}
\affiliation{Center for Studies in Statistical Mechanics and Complex Systems,
\\ The University of Texas at Austin, Austin, TX 78712 USA }

\author{Sungyun Kim}
\affiliation{Max-Planck Institute for Physics of Complex Systems
Noethnizter Str. 38, 01187 Dresden Germany}

\begin{abstract}
It is shown that in a double-cavity, two-dimensional electron waveguide, the interplay between quasi-bound states of each  cavity leads to the appearance of bound states in continuum for certain distances between the cavities. These bound states may be used to trap electrons in de-localized states distributed in both cavities.
\end{abstract}

\maketitle
%

%

Ever since von Neumann and Wigner \cite{wigner} proposed that  certain 
type of oscillating
 attractive potentials  could produce isolated bound states with energies
within the continuum \cite{stillinger, gazdy},
a number of studies have been reported the presence of the ``bound states in 
continuum" (BIC) that can exist above the continuum minimum. 
Fonda and Newton discussed BIC in a system of two 
coupled square-well 
potentials using resonance scattering theory \cite{fonda}. Trajectories of
the poles of this system have been studied in Refs.  \cite{vanroose1, vanroose2}.
Positive energy bound states in super-lattice structures with a single impurity 
potential \cite{nature} or a single defect stub \cite{deo} have been reported.

\begin{figure}
\psfig{file=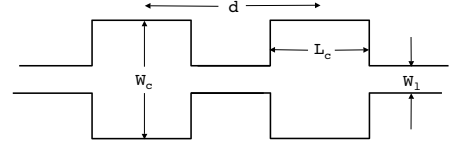, angle=270,width=3in}
\caption{Double cavity electron waveguide}
\label{wguide}
\end{figure}

In Ref. \cite{gonzalo} the existence of BIC was demonstrated for a  model of two atoms in one-dimension. The BIC were formed by the exchange of resonant photons. A physically analogous effect occurs in a two-dimensional, double cavity waveguide shown in Fig.  \ref{wguide}.
The purpose of the present study is to show the the existence of BIC in this waveguide configuration.  Electron waveguides can be formed at a ${\rm GaAs/AlGaAs}$ interface \cite{datta}. 

The leads of the waveguide in Fig.  \ref{wguide} form an infinite quasi-one-dimensional wire, where electrons have a continuous spectrum of energy.  Due to the lateral confinement in the wire, the spectrum has  a minimum energy that allows propagation along the wire. If there is a single cavity, an electron inside the cavity with energy below the minimum  will form bound states. In contrast, an electron with energy above the minimum will form quasi-bound states with finite life-time, where the electron escapes the cavity through the leads.  Bound states and quasi-bound states are associated with real and complex energy poles of the scattering matrix \cite{na}.

For the double-cavity waveguide, the electron states are much more varied. In this study, we  focus on the interplay between the  quasi-bound states formed in each cavity.  We find  BIC appear for certain distances between the cavities.
These positive energy bound states are quite different from  previous ones since they are formed by the interaction between two quasi-bound states.

\smallskip
{\it Theoretical model.}
We use the single-electron Hamiltonian
\begin{eqnarray}
 H = -\frac{\hbar^2}{2m_e^*}\left(\frac{\partial^2}{\partial x^2} +  \frac{\partial^2}{\partial y^2}\right) 
\end{eqnarray}
with vanishing wavefunctions at the boundaries. The effective mass $m_e^*$ is $m_e^*=0.05 m_e$, where $m_e$ is the mass of the free electron. We assume the two cavities are identical. To analyze this Hamiltonian, we decompose the waveguide into two independent closed  cavities and the lead \cite{Suresh0}. The wavefunctions inside the cavities are denoted by $|m,n\ket_i$, where $i=1,2$ labels the cavities and $m$, $n$ are positive integers representing the  horizontal and vertical wave numbers. The wavefunctions in the leads are denoted by $|k,j\ket$ where $k$ is the horizontal wave number (real) and $j$ the vertical wave number (integer). The  energies of the wavefunctions in  either cavity and the lead are, respectively
\begin{eqnarray}
 E_c(m,n) &=& \frac{\hbar^2}{2m_e^*}\left[\left(\frac{m\pi}{L_c}\right)^2 + \left(\frac{n\pi}{W_c}\right)^2\right] \nonumber\\
 E_l(k,j) &=& \frac{\hbar^2}{2m_e^*} \left[k^2 + \left(\frac{j\pi}{W_l}\right)^2\right]
\end{eqnarray}
 The minimum energy for propagation along the lead is $E_l(0,1)$. We consider an electron with low energy narrowly centered around
\begin{eqnarray}
 E_c^0 = E_c(m_0,n_0)
    \EQN{xi0}
\end{eqnarray}
We assume that $ E_l(0,1)<E_c^0 <E_l(0,2)$.
The electron may propagate through the first mode of the lead, but not through the higher ($j>1$) modes. In our theoretical model,  we neglect the $j>1$ modes, which are evanescent, and keep only the $j=1$ mode. Henceforth we omit the $j=1$ index, e.g., $E_l(k)=E_l(k,1)$.

We will rewrite the Hamiltonian using the cavity and lead basis states. With $|i\ket = |m_0,n_0\ket_i$, we define the cavity and lead projectors
\begin{eqnarray}
P_c = \sum_{i=1}^2 |i\ket\bra i|, \quad P_l = \int_{-\infty}^\infty dk\, |k\ket\bra k|
\end{eqnarray}
To make the basis states orthogonal, we introduce the  modified lead states 
\begin{eqnarray}
|\psi_{k}\ket &=&  |k \ket -   R P_c (P_c R  P_c)^{-1} | k \ket, \nonumber\\
R &=& P_l \frac{1}{E_l(k)- H + i0} P_l
\end{eqnarray}
that satisfy $\bra i|\psi_{k}\ket = 0$ and $\bra \psi_{k'}|H| \psi_{k}\ket = E_l(k) \delta(k-k')$. The   following approximate Hamiltonian is obtained 
\begin{eqnarray}
 H &\approx& E^0_c \left[ |1\ket \bra 1| +   |2\ket \bra 2| \right]  
  +   \int_{-\infty}^\infty dk  E_l(k) |\psi_{k}\ket \bra \psi_{k}| \nonumber\\
  &+ & \left[ \int_{-\infty}^\infty dk \sum_{i=1}^2 V_i(k) |i\ket \bra \psi_{k}|
   + \rm{H.c.} \right] 
 \EQN{HWG}
\end{eqnarray}
The terms $V_i (k) = \bra i|H|\psi_{k}\ket$ represent the amplitude of a
transition of the electron from the lead to the
cavities or vice versa. For cavities centered at $x=x_1$ and
$x=x_2$,  they have the form 
\begin{eqnarray}
 V_{1,2}(k) =  v_k e^{i k x_{1,2}} + u_k e^{i k x_{2,1}} 
  \EQN{V12g}
\end{eqnarray}
To construct eigenstates of the  Hamiltonian (\ref{HWG}) we start with the  symmetric and anti-symmetric states 
\begin{eqnarray}
 \label{symm}
 |\pm\ket = \left(|1\ket \pm |2\ket\right)/\sqrt{2}
      \EQN{sadef}
\end{eqnarray}
From these we obtain  eigenstates with complex energy eigenvalues  $z^0_\pm$ \cite{gonzalo}, which are poles of the S-matrix. For $t>0$ we take the eigenvalues with negative imaginary part. They are solutions the integral equation 
 \beqa
 z^0_\pm= E_c^0 + 2 \int_0^\infty dk \frac{ |u_k \pm v_k|^2}{(z^0_\pm -E_l(k))^+}
  (1 \pm \cos k d) 
  \EQN{z0j}
  \eeqa
closest to the real axis.  Here $d=|x_2-x_1|$ is the distance between the cavities. The $+$ superscript means analytic continuation from the upper to the lower half-plane of $z^0_\pm$.  As the distance $d$ is varied, the poles $z^0_\pm$ move in  the complex plane. For certain distances $d_{\pm}$
the imaginary part of $z^0_\pm$ vanishes  \cite{gonzalo}.  This happens when $1 \pm \cos k d_{\pm} = 0$ for $E_l(k) = z^0_\pm$. These conditions give  
 \beqa
  d_\pm = \frac{2n + (1\pm 1)/2}{\sqrt{2 m_e^*(z^0_\pm- E_l(0))}} \pi \hbar
  \EQN{WGg2'}
  \eeqa
with $n$  integer. Replacing $d=d_{\pm}$ in \Ep{z0j} we obtain real solutions for $ z^0_\pm$, which suggests the existence of BIC.   In the following  we verify  this using  a  more accurate description of the electron waveguide. 
%
%
%
%

\smallskip
{\it Computational results.}
We have computed the exact energy eigenstates for the lowest propagating mode
in the double-cavity waveguide as a function of energy and the distance between 
the two cavities
using the boundary integral method \cite{frohne,na}. 
The exact eigenstates are built out of local propagating and evanescent modes
in the leads and cavities and are composed of incoming, $\psi^{-} (x,y)$, and
outgoing, $\psi^{+} (x,y)$ states,
\begin{equation}
\psi(x,y)=\psi^{-} (x,y)+ S(E) \psi^{+} (x,y),
\end{equation}
where the scattering amplitude, $S(E)$, is composed of reflection and transmission
coefficients. 
The unit of the length is the width of the lead, $W_l=1$, which corresponds
to $100 \AA$. 
The width and length of each cavity is $W_c =2$ and $L_c=2$, respectively.

\begin{figure}
 \psfig{file=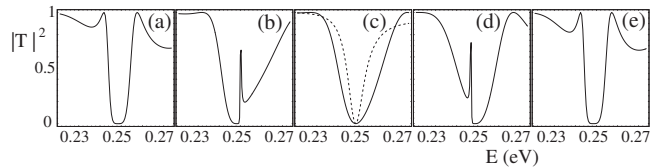, width=3.5in}
\caption{Transmission probability, $|T|^2$, versus energy, $E$, for 
the double-cavity waveguide with $W_c=2$, $L_c=2$, $W_l=1$, and (a) $d=5.25$, (b) $d=5.5$,
(c) $d=5.60$, (d) $d=5.70$, and (e) $d=5.90$.  
The dashed line in (c) is the transmission probability for the single-cavity 
waveguide. }
\label{trans}
\end{figure}

We have computed the transmission probability, $|T|^2$, for the lowest 
propagating mode
as a function of energy for the double-cavity waveguide in Fig. \ref{trans}. 
%
%
Transmission zeros in a given channel have been related to Fano type resonances
between the continuum of that channel and bound states of the closed cavity of higher energy \cite{fano,shao}. The strength of the coupling between the continuum state and the bound state  determines the decay rate and the width of the 
transmission profile \cite{na, itoh}.

As we vary the distance between the two cavities, the transmission profiles
change near the resonance energy region $E=0.25 eV$.
There are sharp peaks in the transmission profiles on either side of the 
resonance energy regions in Figs. \ref{trans} (b) and (d).
There is a broad transmission profile with $d=5.60$ in Fig. \ref{trans} (c). 
For comparison, the transmission probability for the single-cavity waveguide is 
shown in Fig. \ref{trans} (c) as a dotted line. 
The transmission zero profile shows a wider dip for the double-cavity waveguide 
compared with that of the single-cavity waveguide.
We will see later that the BIC is formed when the distance
between the two cavities is $d=5.60$.
These features are associated with the presence of the double poles in the complex energy plane
and affect the dynamics of the electron in the double-cavity waveguide.



The transmission zeros in Fig. \ref{trans} are associated with the poles of 
$T(E)$ in the complex energy plane.
Transmission amplitude in the complex energy plane has a branch cut starting from the lower edge of the continuum and extending 
along the positive energy axis and has poles at energy values $E_R - i \gamma$.
These poles give rise to the transmission zeros on the positive real axis \cite{na,na1}.

\begin{figure} 
 \psfig{file=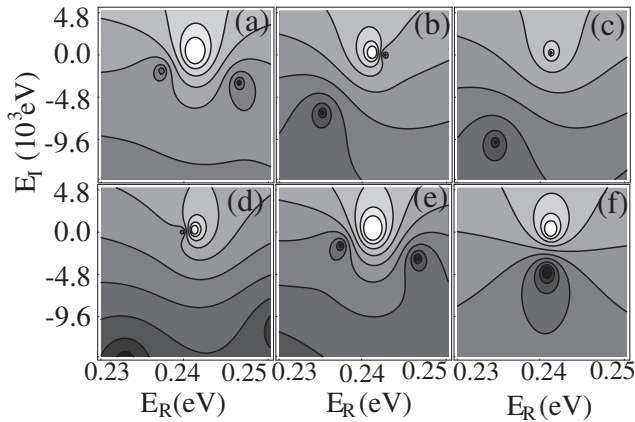, width=3.5in}
\caption{Transmission amplitude, $T(E)$, in the complex energy plane,
$E=E_R + i E_I $, for 
the double-cavity waveguide with $W_c=2$, $L_c=2$, $W_l=1$, and (a) $d=5.25$, (b) $d=5.5$,
(c) $d=5.60$, (d) $d=5.70$, and (e) $d=5.90$.  
The transmission amplitude in the complex energy plane for 
the-single cavity waveguide is in (f).
The dark regions indicate the positions of the poles and the bright region 
indicates the position of the transmission zero.
}
\label{pole_5}
\end{figure}

\begin{figure}
 \psfig{file=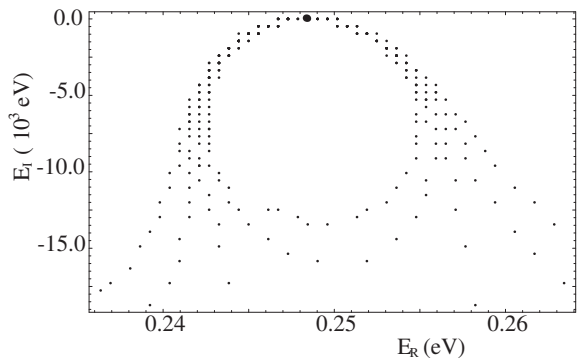,  width=3in}
\caption{Distribution of the pole pairs in the complex energy
plane as we vary the distance between the two cavities.
The dark point on the real axis indicates the energy of the 
transmission zero. 
}
\label{pole_c}
\end{figure}

The locations of the poles in the complex energy plane for the double-cavity waveguide
are shown in Figure \ref{pole_5}. 
%
%
As we increase the distance between the two cavities, the pole located on the 
right-hand side in Fig. \ref{pole_5} (a) approaches the real axis. 
The pole disappears into the real axis and produces a zero value of $\gamma$ 
for the distance, $d=5.60$, in Fig. \ref{pole_5} (c).
This implies the formation of a BIC with infinite life-time.
The pole on the left-hand side in Fig. \ref{pole_5} (a)  moves away from the 
real axis and attains large decay rate  $\gamma$, accordingly. 
As the distance between the cavities is further increased, the pole on the
real axis in Fig. \ref{pole_5} (c) emerges out of the transmission zero and 
recedes from the real axis.
If we continue to increase $d$, the pole on the right-hand side in Fig. \ref{pole_5} (e) approaches  the real  axis and makes a transit through the transmission zero at $d=6.26$ (not shown). 
BIC states regularly reappear as the distance between 
the two cavities is varied. This agrees with the existence of real solutions of  Eq. (\ref{z0j}) with \Ep{WGg2'} for different values of $n$. We have found BIC states for distances of up to $d\approx 50$, which means that these states can be quite de-localized.

In a separate calculation, the wavefunctions inside the two cavities  
for the real energy states associated with the right-hand side and left-hand
side poles in Fig. \ref{pole_5} (a) show symmetric and 
anti-symmetric structures, respectively.
These wavefunctions  are similar to the symmetric and anti-symmetric combinations 
of the eigenstates of the two closed cavities,
\begin{eqnarray}\label{psipm}
\psi_{\pm}(x,y)&=&\frac{1}{\sqrt{2}}\left({\rm sin}(2 \pi (x-x_1)/L_c) {\rm cos}(3 \pi y/W_c) \chi_1 \right.\nonumber\\
           &\pm& \left. {\rm sin}(2 \pi (x-x_2)/L_c) {\rm cos}(3 \pi y/W_c) \chi_2\right)
\label{initial_state}
\end{eqnarray}
where $\chi_{i}=1$ inside cavity $i$ and  $\chi_{i}=0$ outside. Eq. (\ref{psipm}) corresponds to Eq. (\ref{symm}). 

Each complex pole keeps its symmetric or anti-symmetric feature as it migrates throughout the complex energy
plane, as we vary the distance between the two cavities. 
Thus the BIC sstates of the electron appearing at $d=5.6$ and $d=6.26$  are symmetric and anti-symmetric states, respectively. Solutions of Eq. (\ref{z0j}) for $n=4$ and $n=5$ give $d_+=6$ and $d_-=6.67$, respectively, which are in fair agreement with the computational results.

In Figure \ref{pole_c}, we show the distribution of the pole pairs in the 
complex energy plane as we vary the distance between the two cavities.
Each of the poles with either symmetric or anti-symmetric identity
makes a counter-clockwise circulation that passes through the transmission
zero on the real axis.
The dark point on the real axis indicates the energy of the
transmission zero.  
It has been known that the double or multiple poles induced by a laser in an 
atom coalesce to form an exceptional point or repel each other to form an avoided 
crossing \cite{letinne, kylstra, rotter}.
In contrast to these works, the poles of the
double-cavity waveguide make a circular motion and do not approach or repel each
other.


One of the important features of an electron waveguide 
is that the energy spectrum of the electrons is continuous. This means that the electron waveguides fall in the class of unstable systems. An electron in the cavity with wavelength smaller than the critical wavelength for propagation along the lead, will ultimately decay (escape through the leads).   
However, in our study, we have shown that some of the states of an electron
in a double-cavity waveguide can prevent  the state from decaying and
trap the electron inside the two cavities with infinite life-time, forming BIC.

The survival probabilities of an electron placed in the waveguide cavities can
be calculated using the scattering states of the double-cavity waveguide.
The survival probability can be written as $P_{\psi}(t)=|A_{\psi}(t)|^2$,
where $A_{\psi}(t)$ is the survival amplitude,
\begin{equation}
A_{\psi}(t)=\langle \psi | {\rm e}^{-i\hat H t/\hbar} |\psi\rangle 
=\int_0^\infty dE |\langle\psi|E\rangle|^2 {\rm e}^{-i E t/\hbar}.    
\label{surv_p}
\end{equation}
We choose as initial state
$|\psi\ket$, the symmetric or anti-symmetric combinations of the eigenstates 
of the two closed cavities in Eq. (\ref{initial_state}), with no probability amplitude outside the cavities.
In order to solve Eq. (\ref{surv_p}) numerically, we  discretize 
the energy  eigenstates, $|E\rangle$, residing in the continuum. 
%
%
\begin{figure}
\psfig{file=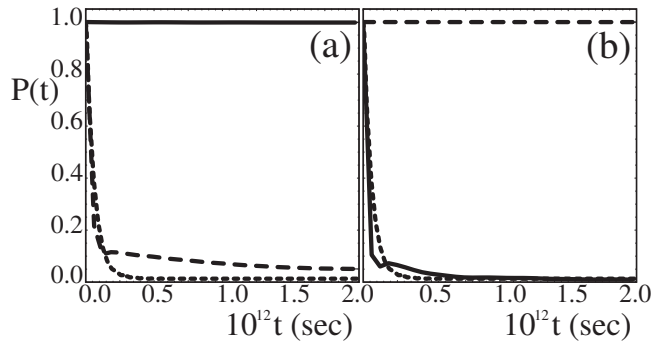, width=3.5in}
\caption{ The survival probability, $P(t)=|A(t)|^2$,
versus time, $t$ for the states prepared symmetrically or anti-symmetrically
inside the two cavities at $t=0$.
The distances between the two cavities are (a) $d=5.60$ and (b) $d=6.26$.
The dotted line shows the survival probability of a state in the  single-cavity
 waveguide in (a) and (b). The solid line displays the survival probability of a
state prepared symmetrically and the dashed line displays the survival
probability of a state prepared anti-symmetrically.}
\label{survival_prob}
\end{figure}

In Fig. \ref{survival_prob}, we plot the survival probability, $P(t)=|A(t)|^2$,
versus time, $t$ for the states prepared symmetrically or anti-symmetrically
inside the two cavities at $t=0$. 
The distances between the two cavities are $d=5.60$ and $d=6.26$  
for Figs. \ref{survival_prob} (a) and (b), respectively.
%
With this arrangement, one of the complex poles has a disappearing imaginary part, 
$\gamma\rightarrow 0$, and gives rise to an infinite life-time. 
The results show that the symmetrically (anti-symmetrically) arranged state does 
not decay over time for the cases with $d=5.60$ ($6.26$). 
On the other hand, the anti-symmetric (symmetric) state decays quickly. 
The dotted line in Fig. \ref{survival_prob} shows the survival probability of 
the state, $\psi={\rm sin}(2 \pi x/L_c) {\rm cos}(3 \pi y/D_c)$, in the single-cavity waveguide.

%


%

In conclusion, we have proved that there can be BIC states with 
specially arranged geometry in double-cavity electron waveguides. 
These maybe used as a quantum information storing device. Pairs of electrons with opposite spins can form de-localized, entangled  states.
The existence of BIC states may be verified experimentally using actual electron waveguides. Alternative experimental setups are electromagnetic waveguides, which are described by a  similar model, and semiconductor arrays of quantum wells \cite{Tomio}.
%
%

We   thank   Professors T. Petrosky,  L. Reichl,  R. Walser and P. Valanju for  helpful comments and suggestions.  We acknowledge the Engineering Research Program of the Office of Basic Energy
Sciences at  the U.S. Department of Energy, Grant No DE-FG03-94ER14465  and U.S. Navy Office of Naval Research, Grant No. N00014-03-1-0639 for partial support of this work.

%
%



\end{document}